\documentstyle[12pt]{article}
\newcommand{\Cslash}{\not \!\! C}
\newcommand{\Dslash}{\not \!\! D}

\newcommand{\sslash}{\not \!\! s}

\newcommand{\f}[2]{\frac{#1}{#2}}

\newcommand{\lb}[0]{\left[}
\newcommand{\rb}[0]{\right]}
\newcommand{\lc}[0]{\left\{}
\newcommand{\rc}[0]{\right\}}
\def\hatmu{\hat{\mu}}
\begin{document}

\begin{flushright}{UT-926}
\end{flushright}

\begin{center}
{\large{\bf Locality Properties of a New Class of Lattice Dirac 
Operators}}
\end{center}
\vskip .5 truecm
\centerline{\bf Kazuo Fujikawa and Masato Ishibashi}
\vskip .4 truecm
\centerline {\it Department of Physics,University of Tokyo}
\centerline {\it Bunkyo-ku,Tokyo 113,Japan}
\vskip 0.5 truecm

\makeatletter
\@addtoreset{equation}{section}
\def\theequation{\thesection.\arabic{equation}}
\makeatother

\begin{abstract}
A new class of lattice Dirac operators $D$ which satisfy the 
index 
theorem  have been recently proposed on the basis of the 
algebraic relation
$\gamma_{5}(\gamma_{5}D) + (\gamma_{5}D)\gamma_{5} =
2a^{2k+1}(\gamma_{5}D)^{2k+2}$. Here $k$ stands for a 
non-negative integer and $k=0$ corresponds to the ordinary
Ginsparg-Wilson relation. We analyze the locality properties 
of Dirac operators which solve the above algebraic relation.
 We first show that the free fermion 
operator is analytic in the entire Brillouin zone for a suitable
choice of parameters $m_{0}$ and $r$, and there exists 
a well-defined ``mass gap'' in 
momentum space, which in 
turn leads to the exponential decay of the operator in 
coordinate space for any finite $k$. This mass gap in the free 
fermion operator 
suggests that the operator is local for sufficiently weak 
background gauge fields. We in fact establish a finite  
locality 
domain of gauge field strength for 
$\Gamma_{5}=\gamma_{5}-(a\gamma_{5}D)^{2k+1}$ 
for any finite $k$, which is 
sufficient for the cohomological analyses of 
 chiral gauge theory.
We also present a crude estimate of the localization length
defined by an exponential decay of the Dirac operator, which 
turns out to be  much shorter than the one given by the general 
Legendre expansion.
\end{abstract}

\section{Introduction}

We have recently witnessed a remarkable progress in the treatment 
of lattice fermions[1]-[4]. The first breakthrough may be traced 
to the domain-wall fermion [5][6], which was followed by the
overlap fermion[7]. See also related works in [8]. It is 
well known that the overlap fermion[2], which developed 
independently of the Ginsparg-Wilson relation[1], satisfies the 
simplest version of the Ginsparg-Wilson relation. The 
recognition that the fermion operator, which satisfies the 
simplest version of the Ginsparg-Wilson relation, gives rise to 
 the index theorem on the 
lattice[3] and thus modified but exact lattice chiral 
symmetry[4], was crucial in the recent developments. The locality 
properties of the Neuberger's overlap operator have also been  
established [9][10]; the operator is not ultra-local[11] but 
exponentially local, which is considered to be sufficient to 
ensure locality in the continuum limit with the lattice 
spacing $a\rightarrow 0$.  

In the mean time, a new class of lattice Dirac operators $D$ 
have been  proposed on the basis of the algebraic 
relation[12]
\begin{equation}
\gamma_{5}(\gamma_{5}D)+(\gamma_{5}D)\gamma_{5}=2a^{2k+1}
(\gamma_{5}D)^{2k+2}
\end{equation}
where $k$ stands for a non-negative integer, and $k=0$ corresponds
to the ordinary Ginsparg-Wilson relation[4] for which an explicit 
example of the operator free of species doubling is known [2]. 
It has been shown in [12] that we can  
construct the lattice Dirac operator, which is free of species 
doublers and satisfies a lattice version of index theorem[3][4], 
for any finite $k$. An explicit calculation of the chiral 
anomaly for these operators has also been 
performed[13]. Here  $\gamma_{5}$ is a 
hermitian chiral Dirac matrix and $\gamma_{5}D$ is hermitian. 
In a sense, this new class of fermion operators are regarded
as the first generation of lattice fermion operators which are 
directly motivated by the Ginsparg-Wilson relation. In contrast 
to the most general form of the Ginsparg-Wilson relation [1], our
algebra (1.1) allows an explicit solution which satisfies 
the index theorem: Our explicit solutions illustrate the features
of possible solutions of the general Ginsparg-Wilson relation.

A salient feature of Dirac operators corresponding to larger
$k$ is that the chiral symmetry breaking term becomes more
irrelevant in the sense of Wilsonian renormalization group.
For $H=a\gamma_{5}D$ in the near continuum 
configurations we have, for example,  
\begin{eqnarray}
H&\simeq&\gamma_{5}ai\Dslash+\gamma_{5}(\gamma_{5}ai\Dslash)^{2}
\ \ for\ \ k=0,\nonumber\\
H&\simeq&\gamma_{5}ai\Dslash+\gamma_{5}(\gamma_{5}ai\Dslash)^{4}
\ \ for\ \ k=1
\end{eqnarray}
respectively. The first terms in these expressions stand for the 
leading terms in chiral symmetric terms, and the second terms in
these expressions stand for the leading terms in chiral symmetry
breaking terms, respectively. This shows that one can improve the
 chiral symmetry for larger $k$, though the operator spreads over 
more lattice points for larger $k$.
As another manifestation
of this property, the spectrum of the 
operators with $k>0$ is closer to that of the continuum operator 
in the sense that the small eigenvalues of $D$ accumulate along 
the imaginary axis ( which is a result of taking a $2k+1$-th 
root), compared to the standard overlap operator  
for which the eigenvalues of $D$ draw a perfect circle in the 
complex eigenvalue plane.

In this paper, we analyze the locality properties of this 
general class of 
lattice Dirac opeartors. As for the overlap Dirac operator[2], 
which corresponds to $k=0$, the very
detailed analyses of locality properties have been performed by
Hernandez, Jansen and L\"{u}scher[9], and Neuberger[10].
A similar locality analysis of the domain wall fermion has 
been given by Kikukawa [14]. 
We establish the locality properties of the general class of 
operators (1.1) for all finite $k$, which are sufficient for 
the cohomological analyses of chiral gauge theory [15][16], 
for example.

\section{A Brief Summary of the Model and Notation}

The explicit construction of the operator, which satisfies 
the relation (1.1),  proceeds by 
first defining 
\begin{equation}
H_{(2k+1)}\equiv(\gamma_{5}aD)^{2k+1}=\frac{1}{2}\gamma_{5}
[1+D_{W}^{(2k+1)}\frac{1}
{\sqrt{(D_{W}^{(2k+1)})^{\dag}D_{W}^{(2k+1)}}}].
\end{equation}
The operator $D_{W}^{(2k+1)}$ is in turn expressed as a
generalization of the ordinary Wilson Dirac operator
as 
\begin{equation}
D_{W}^{(2k+1)}=i(\Cslash)^{2k+1}+(B)^{2k+1}
-(\frac{m_{0}}{a})^{2k+1}.
\end{equation}  

The ordinary Wilson Dirac operator $D_{W}$, which corresponds to 
$D_{W}^{(1)}$, is given by
\begin{eqnarray}
D_{W}(x,y)&\equiv&i\gamma^{\mu}C_{\mu}(x,y)+B(x,y)-
\frac{1}{a}m_{0}\delta_{x,y},\nonumber\\
C_{\mu}(x,y)&=&\frac{1}{2a}[\delta_{x+\hat{\mu} a,y}
U_{\mu}
(y)-\delta_{x,y+\hat{\mu} a}U^{\dagger}_{\mu}(x)],
\nonumber\\
B(x,y)&=&\frac{r}{2a}\sum_{\mu}[2\delta_{x,y}-
\delta_{y+\hat{\mu} a,x}U_{\mu}^{\dagger}(x)
-\delta_{y,x+\hat{\mu} a}U_{\mu}(y)],
\nonumber\\
U_{\mu}(y)&=& \exp [iagA_{\mu}(y)],
\end{eqnarray}
where we added a constant mass term to $D_{W}$.  Our 
matrix convention is that $\gamma^{\mu}$ are anti-hermitian, 
$(\gamma^{\mu})^{\dagger} = - \gamma^{\mu}$, and thus 
$\Cslash\equiv \gamma^{\mu}C_{\mu}(n,m)$ is hermitian
\begin{equation}
\Cslash ^{\dagger} = \Cslash.
\end{equation}
To avoid the appearance of species doublers, we need to satisfy
\begin{equation}
2r>m_{0}>0
\end{equation}
and the choice $2m^{2k+1}_{0}=1$ simplifies the normalization
of the operaror $D$ in the continuum limit as in (1.2). 

By defining hermitian $H\equiv a\gamma_{5}D$, the algebra (1.1) 
is written as 
\begin{equation}
\gamma_{5}H+H\gamma_{5}=2H^{2k+2}
\end{equation}
or equivalently as a set of equations
\begin{eqnarray}
&&\gamma_{5}H^{2k+1}+H^{2k+1}\gamma_{5}=2H^{4k+2},\nonumber\\
&&H^{2}\gamma_{5}-\gamma_{5}H^{2}=0.
\end{eqnarray}
The first of these relations, when regarded as an equation for
$H^{2k+1}$,  is solved by $H_{(2k+1)}$ in (2.1).

The solution of the original algebra (2.6) is then given by
\begin{equation}
H=(H_{(2k+1)})^{1/(2k+1)}
\end{equation}
in the representation where $H_{(2k+1)}$ is diagonal and the 
value of $H$ is chosen as real. This construction means 
$\det H=(\det H_{(2k+1)})^{1/(2k+1)}$. The case $k=0$ is 
reduced to the overlap Dirac operator.

When one defines
\begin{equation}
\Gamma_{5}\equiv \gamma_{5}-H_{(2k+1)}
\end{equation}
the defining algebra is written as
\begin{equation}
\Gamma_{5}H+H\Gamma_{5}=0.
\end{equation}
The (fermionic) index is then given by
\begin{equation}
 Tr\Gamma_{5}=n_{+}-n_{-}
\end{equation}
where $n_{\pm}$ stand for the the number of zero modes 
of 
\begin{equation}
H\varphi_{n}=0
\end{equation}
with $\gamma_{5}\varphi_{n}=\pm\varphi_{n}$, respectively.
In the smooth continuum limit, one can establish the index 
theorem
\begin{equation}
n_{+}-n_{-}=\int d^{4}x\frac{g^{2}}{32\pi^{2}}tr
\epsilon^{\mu\nu\alpha\beta}F_{\mu\nu}F_{\alpha\beta}
\end{equation}
for all (finite) $k$. The right-hand side of this relation
stands for the Pontryagin (or instanton) number. 
See Ref.[12] for further details.

To analyze the locality, one thus has to examine both of 
\begin{equation}
H_{(2k+1)}=H^{2k+1},
\end{equation}
which appears in $\Gamma_{5}$, and $H$ itself. For the free 
fermion case, where one can write
the operator $H$ explicitly, it turns out that the localization 
length is essentially the same for both of $H$ and $H_{(2k+1)}$.
We later argue that this property holds for operators $H$ and 
$H_{(2k+1)}$ with non-trivial gauge fields also.

\section{Mass Gap  for Free Fermion $H^{(2k+1)}_{W}$ }
 
For vanishing gauge fields, we have
\begin{eqnarray}
\sum_{y}D_{W}(x,y)e^{iky}&=&
[-\sum_{\mu}\gamma^{\mu}\frac{\sin ak_{\mu}}{a} + 
\frac{r}{a}\sum_{\mu}(1 - \cos ak_{\mu}) - \frac{m_{0}}{a}]
e^{ikx}\nonumber\\
&\equiv&D_{W}(ak_{\mu})e^{ikx}
\end{eqnarray}
and thus 
\begin{eqnarray}
D_{W}(x,y)&=&\int \frac{d^{4}k}{(2\pi)^{4}}D_{W}(ak_{\mu})
e^{ik(x-y)}\nonumber\\
&=&\frac{1}{a^{4}}\int\frac{d^{4}p}{(2\pi)^{4}}D_{W}(p_{\mu})
e^{ip(x-y)/a}
\end{eqnarray}
where we defined
\begin{equation}
p_{\mu}=ak_{\mu}.
\end{equation}
Note that we are working in the infinite lattice space, and thus 
we have a continuous spectrum in the Brillouin zone, which is 
chosen to be 
\begin{equation}
-\frac{\pi}{2}\leq p_{\mu}<\frac{3\pi}{2}
\end{equation}
for all $\mu=1\sim 4$. 

Similarly we can define $D_{W}^{(2k+1)}(p)$, or 
\begin{eqnarray}
\tilde{D}_{W}^{(2k+1)}(p)&\equiv& (a^{2k+1})D_{W}^{(2k+1)}(p)
\nonumber\\
&=&i[i\sum_{\mu}\gamma^{\mu}\sin p_{\mu}]^{2k+1}
+[r\sum_{\mu}(1-\cos p_{\mu})]^{2k+1}\nonumber\\
&&-(m_{0})^{2k+1}.
\end{eqnarray}
We then have
\begin{eqnarray}
H_{(2k+1)}(p_{\mu})&\equiv&(\gamma_{5}aD(p_{\mu}))^{2k+1}\nonumber
\\
&=&\frac{1}{2}\gamma_{5}[1+\tilde{D}_{W}^{(2k+1)}(p_{\mu})\frac{1}
{\sqrt{(\tilde{D}_{W}^{(2k+1)}(p_{\mu}))^{\dag}
\tilde{D}_{W}^{(2k+1)}(p_{\mu})}}].
\end{eqnarray}
The general operator in the coordinate representation is 
given by
\begin{equation}
H_{(2k+1)}(x,y)=\frac{1}{a^{4}}\int\frac{d^{4}p}{(2\pi)^{4}}
H_{(2k+1)}(p_{\mu})e^{ip(x-y)/a}.
\end{equation}

In the denominator of the general class of Dirac operators
$H_{(2k+1)}(p_{\mu})$ , we thus have
\begin{eqnarray}
F_{(k)}&\equiv&(\tilde{D}_{W}^{(2k+1)}(p_{\mu}))^{\dag}
\tilde{D}_{W}^{(2k+1)}(p_{\mu})\nonumber\\
&=& [s^{2}]^{2k+1}
+\{[\sum_{\mu}(1-c_{\mu})]^{2k+1}-(m_{0})^{2k+1}\}^{2}\nonumber\\
&=&[\sum_{\mu}(1-c^{2}_{\mu})]^{2k+1}
+\{[\sum_{\mu}(1-c_{\mu})]^{2k+1}-(m_{0})^{2k+1}\}^{2} 
\end{eqnarray}
where we defined the variables
\begin{eqnarray}
&&p_{\mu}= ak_{\mu},\nonumber\\
&&\sslash=\sum_{\mu}\gamma^{\mu}\sin p_{\mu},\nonumber\\
&&s^{2}=\sum_{\mu}(\sin p_{\mu})^{2},\nonumber\\
&&s_{\mu}=\sin p_{\mu},\nonumber\\
&&c_{\mu}=\cos p_{\mu}
\end{eqnarray}
and we use the value of the Wilson parameter
\begin{equation}
r=1
\end{equation}
throughout  the present paper.
We also set
\begin{equation}
m^{2k+1}_{0}= 1.
\end{equation}

\subsection{Minimum of $F_{(k)}$}

To analyze the ``mass gap'' of $F_{(k)}$, we examine
\begin{eqnarray}
F_{(k)}&=&[\sum_{\mu}(1-c^{2}_{\mu})]^{2k+1}
+\{[\sum_{\mu}(1-c_{\mu})]^{2k+1}-1 \}^{2}\nonumber\\
&=&[\sum_{\mu}(1-c^{2}_{\mu})]^{2k+1}
+[\sum_{\mu}(1-c_{\mu})^{2}]^{2k+1}-
2[\sum_{\mu}(1-c_{\mu})]^{2k+1}\nonumber\\
&&+[\sum_{\mu}(1-c_{\mu})]^{2(2k+1)}
-[\sum_{\mu}(1-c_{\mu})^{2}]^{2k+1}+1.
\end{eqnarray}
We first observe that 
\begin{equation}
[\sum_{\mu}(1-c_{\mu})]^{2(2k+1)}
-[\sum_{\mu}(1-c_{\mu})^{2}]^{2k+1}\geq 0
\end{equation}
by noting 
\begin{equation}
[\sum_{\mu}(1-c_{\mu})]^{2}\geq [\sum_{\mu}(1-c_{\mu})^{2}]\geq 0
\end{equation}
where the equality holds only for $c_{\mu}=1$ for 
$d-1$ indices $\mu$ in $d\geq 2$.

We next define 
\begin{eqnarray}
&&B\equiv \sum_{\mu}(1-c_{\mu})\geq 0,\nonumber\\
&&A+B\equiv \sum_{\mu}(1-c_{\mu})^{2}
\end{eqnarray}
then one can confirm
\begin{equation}
-A+B=-(A+B)+2B=\sum_{\mu}(1-c^{2}_{\mu}).
\end{equation}
We thus have
\begin{eqnarray}
&&[\sum_{\mu}(1-c^{2}_{\mu})]^{2k+1}
+[\sum_{\mu}(1-c_{\mu})^{2}]^{2k+1}-
2[\sum_{\mu}(1-c_{\mu})]^{2k+1}\nonumber\\
&&=(-A+B)^{2k+1}+(A+B)^{2k+1}-2B^{2k+1}\nonumber\\
&&=2\sum_{l=even,\neq 0}\left(
 \begin{array}{c}
  2k+1\\ l\\
 \end{array}\right)A^{l}B^{2k+1-l}\geq 0
\end{eqnarray}
since $B\geq 0$. Here the equality sign holds only for
$B=A=0$ with $c_{\mu}=1$ for all $\mu$, or $A=0$ but 
$B\neq0$ with $\sum_{\mu}c_{\mu}(1-c_{\mu})=0$.

By this way, 
we have established that for any finite $k$
\begin{equation}
F_{(k)}\geq 1
\end{equation}
where the equality holds only for $c_{\mu}=1$ for all $\mu$
or one of $c_{\mu}=0$ and the rests of $c_{\mu}=1$ in the 
space-time dimensions $d\geq 2$.
Namely 
\begin{equation}
||\tilde{H}^{(2k+1)}_{W}||\equiv||\gamma_{5}\tilde{D}_{W}^{(2k+1)}
(p_{\mu})||\geq 1
\end{equation}
for $m_{0}=1$. 
If one combines this relation with
\begin{equation}
\frac{\partial \lambda(m_{0})}{\partial m^{2k+1}_{0}}
=\frac{\partial}{\partial m^{2k+1}_{0}}
(\phi,\tilde{H}^{(2k+1)}_{W}\phi )    
=(\phi,\gamma_{5}\phi )
\end{equation}
where $\lambda(m_{0})$ is the eigenvalue of the hermitian 
operator $\tilde{H}^{(2k+1)}_{W}$, we obtain 
a ``flow inequality''[10]
\begin{equation}
|\frac{\partial \lambda(m_{0})}{\partial m^{2k+1}_{0}}|
=|(\phi,\gamma_{5}\phi )|\leq ||\phi||||\gamma_{5}\phi||= 1.
\end{equation}
We can then establish the  ``mass gap'' of  
 $||\tilde{H}^{(2k+1)}_{W}(m_{0})||$ for other choices 
of $m^{2k+1}_{0}$. 
Following the analysis in Ref.[10], one can establish, 
for example\footnote{In this case, the equality holds only for 
$c_{\mu}=1$ for all $\mu$.}, 
\begin{equation}
||\tilde{H}^{(2k+1)}_{W}||\equiv||\gamma_{5}\tilde{D}_{W}^{(2k+1)}
(p_{\mu})||\geq \frac{1}{2}
\end{equation}
for 
\begin{equation}
2m^{2k+1}_{0}=1
\end{equation}
which simplifies the normalization of the continuum limit 
as in (1.2).

\subsection{Maximum of $F_{(k)}$}

To analyze the maximum of $F_{(k)}$, 
we define
\begin{equation}
F_{(k)}(c)=[1-c^{2}+\Sigma^{(2)}]^{2k+1}
+\{[1-c+\Sigma^{(1)}]^{2k+1}-1\}^{2}
\end{equation}
where we defined $c$ as one of four variables $c_{\mu}$, and
\begin{eqnarray}
\Sigma^{(2)}=\sum^{\prime}_{\mu}(1-c^{2}_{\mu})\geq 0,\nonumber\\
\Sigma^{(1)}=\sum^{\prime}_{\mu}(1-c_{\mu})\geq 0.
\end{eqnarray}
Here, the summation runs over the three indices 
$\mu$ except for the one used for $c=c_{\mu}$.
We then obtain
\begin{equation}
F^{\prime}_{(k)}(c)=-2(2k+1)\{[1-c^{2}+\Sigma^{(2)}]^{2k}c
+\{[1-c+\Sigma^{(1)}]^{2k+1}-1\}
[1-c+\Sigma^{(1)}]^{2k}\}.
\end{equation}

We now examine one special section of species doublers specified
by
\begin{equation}
c_{\mu}\leq 0 \ \ \ for\ \ \ all\ \ \ \mu.
\end{equation}
In this domain, the above formula of  $F^{\prime}_{(k)}(c)$ 
gives rise to 
\begin{eqnarray}
&&-\frac{1}{2(2k+1)}F^{\prime}_{(k)}(c)\nonumber\\
&=&
\{[1-c^{2}+\Sigma^{(2)}]^{2k}c
+\{[1-c+\Sigma^{(1)}]^{2k+1}-1\}
[1-c+\Sigma^{(1)}]^{2k}\}\nonumber\\
&\geq&[1-c+\Sigma^{(1)}]^{2k}\{c
+[1-c+\Sigma^{(1)}]^{2k+1}-1\}\nonumber\\
&\geq&[1-c+\Sigma^{(1)}]^{2k}\{[1-c+\Sigma^{(1)}]^{2k+1}
-2\}\nonumber\\
&\geq&[1-c+\Sigma^{(1)}]^{2k}\{[4]^{2k+1}
-2\} >0.
\end{eqnarray}
Here we used 
\begin{eqnarray}
0\leq 1-c^{2}+\Sigma^{(2)}
&=&(1-c)(1+c)+\sum^{\prime}_{\mu}(1-c_{\mu})(1+c_{\mu})\nonumber\\
&\leq& 1-c+\Sigma^{(1)}
\end{eqnarray}
in the above domain, and thus 
\begin{equation}
[1-c^{2}+\Sigma^{(2)}]c\geq [ 1-c+\Sigma^{(1)}]c.
\end{equation}
This shows that $F^{\prime}_{(k)}(c)<0$ for any choice of 
$c=c_{\mu}$. Thus the maximum of $F_{(k)}$ appears at 
$c_{\mu}=-1$ for all $\mu$ in this domain.

This value is also the absolute maximum of $F_{(k)}$ in the 
entire Brillouin zone. This is because
\begin{eqnarray}
F_{(k)}(c_{1},c_{2},c_{3},c_{4})
&=&[\sum_{\mu}(1-c^{2}_{\mu})]^{2k+1}
+\{[\sum_{\mu}(1-c_{\mu})]^{2k+1}-1\}^{2}\nonumber\\ 
&\geq&F_{(k)}(-c_{1},c_{2},c_{3},c_{4})
\end{eqnarray}
if all $c_{\mu}\leq 0$: One can in fact confirm that 
$F_{(k)}(c_{1},c_{2},c_{3},c_{4})$ is not smaller than the 
function
obtained from it by reversing the signatures of some of $c_{\mu}$.
This means that the maximum of $F_{(k)}$ in the domain
$c_{\mu}\leq 0$ for all $\mu$ is in fact the absolute maximum of
$F_{(k)}$ , which is given by setting all $c_{\mu}=-1$
\begin{equation}
F_{(k)}\leq(8^{2k+1}-1)^{2}.
\end{equation}
The absolute maximum of $F_{(k)}=
(\tilde{D}_{W}^{(2k+1)}(p_{\mu}))^{\dag}
\tilde{D}_{W}^{(2k+1)}(p_{\mu})$ provides the upper bound 
for the eigenvalue and thus the norm of the operator 
$(\tilde{D}_{W}^{(2k+1)})^{\dag}\tilde{D}_{W}^{(2k+1)}$. 

\subsection{Locality of Free $H_{(2k+1)}(x,y)$ }

Following the analysis of Ref.[9], we can then establish 
the locality bound for the free operator $D_{(2k+1)}(x,y)
=\gamma_{5}H_{(2k+1)}(x,y)$ in (2.1) by
\begin{equation}
|\frac{1}
{\sqrt{(\tilde{D}_{W}^{(2k+1)})^{\dag}\tilde{D}_{W}^{(2k+1)}}}|
\leq \frac{\kappa}{1-t}
\exp[-\frac{\theta ||x-y||}{2(2k+1)a}].
\end{equation}
When one denotes the maximum and minimum of $F_{(k)}$
by $v_{max}$ and $v_{min}$, respectively, the parameters are 
defined by
\begin{eqnarray}
&&\cosh\theta=\frac{v_{max}+v_{min}}{v_{max}-v_{min}},\nonumber\\
&&t=e^{-\theta},\nonumber\\
&&\kappa=\sqrt{4t/(v_{max}-v_{min})}.
\end{eqnarray}
For $v_{max}\gg v_{min}$, we obtain
\begin{equation}
\theta\simeq2 \sqrt{\frac{v_{min}}{v_{max}}}
\end{equation}
and the localization length, which is defined by the exponential
decay $\sim\exp[-||x-y||/L]$, is estimated by
\begin{equation}
L\simeq (2k+1)a\sqrt{\frac{v_{max}}{v_{min}}}.
\end{equation}

In the present case
\begin{equation}
\cosh\theta=\frac{(8^{2k+1}-1)^{2}+1}
{(8^{2k+1}-1)^{2}-1}
\end{equation}
One thus obtains
\begin{equation}
\cosh\theta\simeq 1+2\times 8^{-2(2k+1)}
\end{equation}
or
\begin{equation}
\theta\simeq 2\times 8^{-(2k+1)}.
\end{equation}
The localization length $L$ is then estimated by 
\begin{equation}
L\simeq (2k+1)2^{3(2k+1)}a
\end{equation}
or in 2-dimensional case
\begin{equation}
L\simeq (2k+1)2^{2(2k+1)}a.
\end{equation}
We will later show that this localization length is much bigger 
than our estimate on the basis of analyticity, which is closer to
the numerical estimate.

We here present a suggestive interpretation of the above estimate 
of $\cosh\theta$ (3.34). The above formula is written as 
\begin{equation}
\cosh\theta=1+\frac{2v_{min}}{v_{max}-v_{min}}.
\end{equation}
This last relation is re-written as
\begin{equation}
v_{min}+\frac{v_{max}-v_{min}}{\cos\pi-\cos 0}(\cos(0+i\theta)-
\cos 0)=0
\end{equation}
by noting $\cos i\theta=\cosh\theta$. This formula 
is interpreted  that
one is estimating the zero of a generic function $f(\cos\theta)$
of a single variable $\cos\theta$, which has a minimum 
$f(\cos\theta)=v_{min}$ at $\cos 0=1$ and a maximum 
$f(\cos\theta)=v_{max}$ at $\cos\pi=-1$. As a crude 
approximation for such a function,
one may consider a function
\begin{equation}
f(\cos\theta)=v_{min}
+\frac{v_{max}-v_{min}}{\cos\pi-\cos 0}(\cos\theta-
\cos 0)
\end{equation}
which has an average  slope 
\begin{equation}
\frac{v_{max}-v_{min}}{\cos\pi-\cos 0}
\end{equation}
at $\cos 0=1$. One may then look for the position of the zero of 
 $f(\cos\theta)=0$, which is located 
closest to the real axis. This equation corresponds to (3.43)
above. The zero of $f(\cos\theta)$ gives rise to a singularity
in $\sqrt{f(\cos\theta)}$, which in turn gives rise to the 
localization length in an analogous analysis in Section 5
later.  

This estimate of the zero (3.43) may be a reasonable one if one 
knows 
only the values of $v_{max}$ and $v_{min}$ in the entire 
domain $-1\leq \cos\theta\leq 1$ but no 
more information about the 
functional form of $f(\cos\theta)$: But this estimate becomes 
poor if all the derivatives of $f(\cos\theta)$ at 
$\cos\theta=1$ vanish  
$f^{(l)}(\cos\theta=1)=0$ up to a certain integer $l$, 
for example. More about this interpretation will be commented on 
later.

\section{Free Fermion Operator $H=a\gamma_{5}D$}

We first define an explicit form of $H$ in 
the free fermion case.
We start with an ansatz of the general solution as (see also
Ref.[17]) 
\begin{equation}
H=\frac{1}{2}\gamma_{5}[A+B\frac{\sslash}{a}]
\end{equation}
and evaluate $2k+1$-th power. We first note
\begin{equation}
H^{2}=\frac{1}{4}[A-B\frac{\sslash}{a}][A+B\frac{\sslash}{a}]
     =\frac{1}{4}[A^{2}+B^{2}\frac{s^{2}}{a^{2}}]
\end{equation}
which satisfies the necessary condition (2.7) 
\begin{equation}
\gamma_{5}H^{2}-H^{2}\gamma_{5}=0.
\end{equation}
Here, we defined
\begin{eqnarray}
\sslash=\sum_{\mu}\gamma^{\mu}\sin ak_{\mu},\nonumber\\
s^{2}=\sum_{\mu}(\sin ak_{\mu})^{2}\geq 0.
\end{eqnarray}
Our convention is $(\gamma^{\mu})^{\dagger}=-\gamma^{\mu}$.
Our ansatz thus gives 
\begin{equation}
H^{2k+1}=(\frac{1}{2})^{2k+1}[A^{2}+B^{2}\frac{s^{2}}{a^{2}}]^{k}
\gamma_{5}[A+B\frac{\sslash}{a}]
\end{equation}

On the other hand, our explicit construction uses the building 
block 
\begin{equation}
H_{W}=\gamma_{5}\{-(\frac{s^{2}}{a^{2}})^{k}\frac{\sslash}{a}
+[\sum_{\mu}\frac{r}{a}(1-\cos ak_{\mu})]^{2k+1}
-(\frac{m_{0}}{a})^{2k+1} \}
\end{equation}
and 
\begin{equation}
H^{2k+1}=\frac{1}{2}\gamma_{5}[1+\gamma_{5}H_{W}\frac{1}
{\sqrt{H^{2}_{W}}}].
\end{equation}
We now define
\begin{eqnarray}
H^{2}_{W}&=&(\frac{s^{2}}{a^{2}})^{2k+1}
+\{[\sum_{\mu}\frac{r}{a}(1-\cos ak_{\mu})]^{2k+1}
-(\frac{m_{0}}{a})^{2k+1}\}^{2}\nonumber\\
&=&(\frac{s^{2}}{a^{2}})^{2k+1}+M_{k}^{2},\nonumber\\
M_{k}&\equiv&[\sum_{\mu}\frac{r}{a}(1-\cos ak_{\mu})]^{2k+1}
-(\frac{m_{0}}{a})^{2k+1}.
\end{eqnarray}
Then our construction gives 
\begin{equation}
H^{2k+1}=\frac{1}{2}\gamma_{5}\{1
+[M_{k}-(\frac{s^{2}}{a^{2}})^{k}\frac{\sslash}{a}]
\frac{1}{\sqrt{H^{2}_{W}}}\}
\end{equation}

Comparing these two expressions (4.5) and (4.9), we have the 
relations
\begin{eqnarray}
&&(\frac{1}{2})^{2k}[A^{2}+B^{2}\frac{s^{2}}{a^{2}}]^{k}A=
1+\frac{M_{k}}{\sqrt{H^{2}_{W}}},\nonumber\\
&&(\frac{1}{2})^{2k}[A^{2}+B^{2}\frac{s^{2}}{a^{2}}]^{k}B=
-\frac{1}{\sqrt{H^{2}_{W}}}(\frac{s^{2}}{a^{2}})^{k}.
\end{eqnarray}
From this we obtain
\begin{equation}
(\frac{1}{2})^{4k}[A^{2}+B^{2}\frac{s^{2}}{a^{2}}]^{2k+1}
=2[1+\frac{M_{k}}{\sqrt{H^{2}_{W}}}]
\end{equation}
by noting the relation
\begin{equation}
H^{2}_{W}-M_{k}^{2}=(\frac{s^{2}}{a^{2}})^{2k+1},
\end{equation}
and thus
\begin{equation}
[A^{2}+B^{2}\frac{s^{2}}{a^{2}}]^{k}=2^{2k}2^{-\frac{k}{2k+1}}
[1+\frac{M_{k}}{\sqrt{H^{2}_{W}}}]^{\frac{k}{2k+1}}.
\end{equation}
By this way we obtain
\begin{eqnarray}
A&=&2^{\frac{k}{2k+1}}
[1+\frac{M_{k}}{\sqrt{H^{2}_{W}}}]^{\frac{k+1}{2k+1}}
\nonumber\\
&=&2^{\frac{k}{2k+1}}
(\frac{1}{\sqrt{H^{2}_{W}}})^{\frac{k+1}{2k+1}}
(\sqrt{H^{2}_{W}}+M_{k})^{\frac{k+1}{2k+1}},\nonumber\\
B&=&-2^{\frac{k}{2k+1}}\frac{1}{\sqrt{H^{2}_{W}}}
(\frac{s^{2}}{a^{2}})^{k}
[1+\frac{M_{k}}{\sqrt{H^{2}_{W}}}]^{-\frac{k}{2k+1}}\nonumber\\
&=&-2^{\frac{k}{2k+1}}
(\frac{1}{\sqrt{H^{2}_{W}}})^{\frac{k+1}{2k+1}}
(\sqrt{H^{2}_{W}}-M_{k})^{\frac{k}{2k+1}}.
\end{eqnarray}
Namely
\begin{eqnarray}
H&=&\gamma_{5}(\frac{1}{2})^{\frac{k+1}{2k+1}}
(\frac{1}{\sqrt{H^{2}_{W}}})^{\frac{k+1}{2k+1}}
\{(\sqrt{H^{2}_{W}}+M_{k})^{\frac{k+1}{2k+1}}
-(\sqrt{H^{2}_{W}}-M_{k})^{\frac{k}{2k+1}}
\frac{\sslash}{a} \}\nonumber\\
&=&\gamma_{5}(\frac{1}{2})^{\frac{k+1}{2k+1}}
(\frac{1}{\sqrt{F_{(k)}}})^{\frac{k+1}{2k+1}}
\{(\sqrt{F_{(k)}}+\tilde{M}_{k})^{\frac{k+1}{2k+1}}
-(\sqrt{F_{(k)}}-\tilde{M}_{k})^{\frac{k}{2k+1}}
\sslash \}\nonumber\\
&&
\end{eqnarray}
where
\begin{eqnarray}
F_{(k)}&=&(s^{2})^{2k+1}+\{[\sum_{\mu}(1-c_{\mu})]^{2k+1}
-1\}^{2},\nonumber\\
\tilde{M}_{k}&=&[\sum_{\mu}(1-c_{\mu})]^{2k+1}
-1.
\end{eqnarray}
\\

We have established that $F_{(k)}\geq 1$ in Section 3. To 
establish the 
analyticity of $H$ in the entire Brillouin zone,
 we thus have to examine 
\begin{eqnarray}
f_{\pm}&\equiv& \sqrt{F_{(k)}}\pm \tilde{M}_{k}\nonumber\\
&=& \sqrt{(s^{2})^{2k+1}
+\{[\sum_{\mu}(1-c_{\mu})]^{2k+1}
-1\}^{2}}\pm\{[\sum_{\mu}(1-c_{\mu})]^{2k+1}
-1\}.\nonumber\\
&&
\end{eqnarray}
If we find a non-analytic behavior of 
\begin{equation}
f_{+}^{\frac{k+1}{2k+1}},\ \ or\ \ f_{-}^{\frac{k}{2k+1}}
\end{equation}
inside the Brillouin zone, we are in danger to encounter the 
non-locality of our operators $H$ in (4.15). \\

We first note that we can establish  
\begin{equation}
f_{-}>\frac{1}{2}
\end{equation}
in the physical domain by using
\begin{eqnarray}
f_{-}&=& \sqrt{[\sum_{\mu}(1-c^{2}_{\mu})]^{2k+1}
+\{[\sum_{\mu}(1-c_{\mu})]^{2k+1}
-1\}^{2}}-[\sum_{\mu}(1-c_{\mu})]^{2k+1}\nonumber\\
&&+1\nonumber\\
&=&\frac{[\sum_{\mu}(1-c^{2}_{\mu})]^{2k+1}- 
[\sum_{\mu}(1-c_{\mu})]^{2k+1}+\frac{3}{4}}
{\sqrt{[\sum_{\mu}(1-c^{2}_{\mu})]^{2k+1}
+\{[\sum_{\mu}(1-c_{\mu})]^{2k+1}
-1\}^{2}}+ [\sum_{\mu}(1-c_{\mu})]^{2k+1}-\frac{1}{2}}
\nonumber\\
&&+\frac{1}{2}
\nonumber\\
&>&\frac{1}{2}
\end{eqnarray}
since $[\sum_{\mu}(1-c^{2}_{\mu})]^{2k+1}
- [\sum_{\mu}(1-c_{\mu})]^{2k+1}\geq 0 $ in the physical domain. 
We also used the gap relation
$\sqrt{[\sum_{\mu}(1-c^{2}_{\mu})]^{2k+1}
+\{[\sum_{\mu}(1-c_{\mu})]^{2k+1}
-1\}^{2}}\geq 1 $.
\\

We next note the identities with a non-negative integer $k$
\begin{eqnarray}
&&f_{-}^{\frac{k}{2k+1}}\equiv (s^{2})^{k}/f_{+}^{\frac{k}{2k+1}}
=(\sum_{\mu}(1-c^{2}_{\mu}))^{k}/f_{+}^{\frac{k}{2k+1}},
\nonumber\\
&&f_{+}^{\frac{k+1}{2k+1}}\equiv (s^{2})^{k+1}
/f_{-}^{\frac{k+1}{2k+1}}=(\sum_{\mu}(1-c^{2}_{\mu}))^{k+1}
/f_{-}^{\frac{k+1}{2k+1}}.
\end{eqnarray}
This shows that we can establish the analyticity of 
$f^{\frac{k}{2k+1}}_{-}$ in the entire Brillouin zone, 
if we can show that $f_{+}$ is positive and non-vanishing in the 
domains of ``species doublers''.

In a domain of species doublers, for example,  $c_{1}<0$
and other $c^{\prime}_{\mu}>0$, 
we have
\begin{eqnarray}
f_{+}&=&
\sqrt{[1-c^{2}_{1}+\sum^{\prime}_{\mu}(1-c^{2}_{\mu})]^{2k+1}
+\{[1-c_{1}+\sum^{\prime}_{\mu}(1-c_{\mu})]^{2k+1}
-1\}^{2}}\nonumber\\
&&+[1-c_{1}+\sum^{\prime}_{\mu}(1-c_{\mu})]^{2k+1}
-1\nonumber\\
&\geq&
\sqrt{[1-c^{2}_{1}+\sum^{\prime}_{\mu}(1-c^{2}_{\mu})]^{2k+1}
+\{[1-c_{1}+\sum^{\prime}_{\mu}(1-c_{\mu})]^{2k+1}
-1\}^{2}}\nonumber\\
&\geq& 1
\end{eqnarray}
where the equality does not hold simultaneously.
Here $\sum^{\prime}_{\mu}$ stands for the sum over $\mu=2\sim4$. 
We thus have
\begin{equation}
f_{+}>1.
\end{equation}

By repeating a similar analysis of $f_{+}$ for other domains of 
the species doublers, we can establish the positivity of 
$f_{+}$ in all the domains of species doublers and thus the 
analyticity of $f_{-}^{\frac{k}{2k+1}}$ in the entire Brillouin 
zone.
Similarly, we can establish the analyticity of 
$f_{+}^{\frac{k+1}{2k+1}}$ in the entire Brillouin zone, by 
using the above identity (4.21) and the positivity of $f_{-}$ in 
the physical domain.

We have thus established the analyticity of the free $H$ for any 
finite $k$ in the entire Brillouin zone. 

\section{Locality with  Exponential Decay}

We analyze the asymptotic behavior of the operator $D$ by means 
of the Fourier transform of the momentum representation.
We first  examine $f_{-}(c_{\mu})^{\frac{k}{2k+1}}$ 
appearing in $H$ with
\begin{eqnarray}
f_{-}(c_{\mu})&\equiv&\sqrt{(\sum_{\mu}(1-c^{2}_{\mu}))^{2k+1}
+\{[\sum_{\mu}(1-c_{\mu})]^{2k+1}
-1\}^{2}}\nonumber\\
&&-[\sum_{\mu}(1-c_{\mu})]^{2k+1}
+1
\end{eqnarray}
as a function of complex $c_{\mu}$ by fixing the phase 
convention in the physical domain. It can be 
confirmed that $f_{-}$
vanishes only for $c^{2}_{\mu}=1$ for all $\mu$ even if one 
analytically extends
$f_{-}$ as a function of $c_{\mu}$. But at $c_{\mu}=1$ for all
$\mu$,
$f_{-}=2\neq 0$ and thus  $f_{-}(c_{\mu})^{\frac{k}{2k+1}}$ is 
regular in the neighborhood of the domain of physical
species. One can also establish that $f_{+}$ is non-vanishing
in the neighborhood of the domain of species doublers even if 
one extends analytically in $c_{\mu}$. By recalling the 
identities (4.21), 
one can define $f_{-}(c_{\mu})^{\frac{k}{2k+1}}$
in the domain of species doublers by using $f_{+}$.
By this way, one can establish that 
$f_{-}(c_{\mu})^{\frac{k}{2k+1}}$ is analytic in the neighborhood
of the entire Brillouin zone.
To be more precise,  $f_{-}(c_{\mu})^{\frac{k}{2k+1}}$ is analytic
in the strip of complex $p_{\mu}$ plane with 
$c_{\mu}=\cos p_{\mu}$,
\begin{eqnarray}
&&-\frac{\pi}{2}\leq \Re p_{\mu} \leq \frac{3\pi}{2},\nonumber\\
&& -\infty < \Im p_{\mu} < \infty
\end{eqnarray}
{\em except} for the singularities appearing in 
\begin{equation}
\sqrt{F_{k}(c_{\mu})}=\sqrt{(\sum_{\mu}(1-c^{2}_{\mu}))^{2k+1}
+\{[\sum_{\mu}(1-c_{\mu})]^{2k+1}
-1\}^{2}}.
\end{equation}
Similarly, one can show the analyticity of 
$f_{+}(c_{\mu})^{\frac{k+1}{2k+1}}$ in the same domain of 
complex $p_{\mu}$ except for the singularities of 
$\sqrt{F_{k}(c_{\mu})}$.  
One thus establishes that the numerator factors of the free 
fermion operator $H$ in (4.15) do not induce singularities other 
than 
those of  $\sqrt{F_{k}(c_{\mu})}$ even in the complex $c_{\mu}$ 
plane.
 
We thus encounter the singularities of the operator $H$ only at 
the singularities of $\sqrt{F_{k}(c_{\mu})}$, which appears 
in the denominator of $H$,  
in the complex $c_{\mu}$ plane. The existence of the mass gap 
means that those singularities are away from the real axis by 
a {\em finite} distance. It is important to recognize that  the 
possible singularities are located at the same points  for both 
of free $H$ and free $H^{2k+1}=H_{(2k+1)}$, which are controlled
by the behavior of $\sqrt{F_{k}(c_{\mu})}$. Since we have
shown that 
free $H^{2k+1}=H_{(2k+1)}$ is local for any finite $k$ in 
Section 4, we infer the locality of free $H$ for any finite $k$. 
\\

In the following we establish the locality of free $H$ directly 
by examining 
\begin{eqnarray}
D(x^{\mu},0)=\frac{1}{a}\int_{-\pi/2}^{3\pi/2} 
\gamma_{5}H(p_{\mu})
\exp[ip_{\mu}x^{\mu}/a]d^{4}p_{\mu}/a^{4}
\end{eqnarray}
for large positive $x^{1}$. Our metric convention is 
$g_{\mu\nu}=(1,1,1,1)$. 
We replace
\begin{equation}
\sslash\rightarrow ia\Cslash
\end{equation}
in the numerator of $H(p_{\mu})$, and thus this factor does not 
influence the analysis of locality. We set 
$x^{\mu}=0$ for $\mu\neq 1$.

When one lets $x^{1}$ {\em positive} and very large, we shift the 
integration path 
of $p_{1}$ to a line parallel to the real axis in the upper half 
complex plane. We note that the integrand $\gamma_{5}H(p)$
is identical at $p_{1}=-\pi/2+iy$ and $p_{1}=3\pi/2+iy$ for 
a real $y$. Also,  $x^{1}$ assumes the values only on the 
lattice points and thus 
\begin{equation}
\exp[i(3\pi/2+iy)x^{1}/a]=\exp[i(-\pi/2+iy)x^{1}/a].
\end{equation}
The contributions in the line integral along the vertical lines 
at $\Re p_{1}=3\pi/2$ and $\Re p_{1}=-\pi/2$  thus cancel.

The leading behavior in $x^{1}$ is thus determined by the 
integration path which is parallel to the real axis and  first 
touches the singularity 
(i.e., the singularity closest to the real axis) in the complex 
$p_{1}$ plane(See Figure 1). See also Ref.[18].

\begin{figure}
\begin{center}
%WinTpicVersion2.15
\unitlength 0.1in
\begin{picture}(38.80,24.25)(12.14,-26.20)
% VECTOR 2 0 3 0
% 2 1398 2781 5094 2781
% 
\special{pn 8}%
\special{pa 1398 2381}%
\special{pa 5094 2381}%
\special{fp}%
\special{sh 1}%
\special{pa 5094 2381}%
\special{pa 5027 2361}%
\special{pa 5041 2381}%
\special{pa 5027 2401}%
\special{pa 5094 2381}%
\special{fp}%
% STR 2 0 3 0
% 3 1574 2865 1574 2949 5 0
% $-\pi/2$
\put(15.7400,-25.4900){\makebox(0,0){$-\pi/2$}}%
% STR 2 0 3 0
% 3 4262 2865 4262 2949 5 0
% $3\pi/2$
\put(42.6200,-25.4900){\makebox(0,0){$3\pi/2$}}%
% LINE 2 0 3 0
% 2 4262 2781 4262 2277
% 
\special{pn 8}%
\special{pa 4262 2381}%
\special{pa 4262 1877}%
\special{fp}%
% LINE 2 0 3 0
% 2 4262 2285 1566 2285
% 
\special{pn 8}%
\special{pa 4262 1885}%
\special{pa 1566 1885}%
\special{fp}%
% LINE 2 0 3 0
% 2 1566 2285 1566 2781
% 
\special{pn 8}%
\special{pa 1566 1885}%
\special{pa 1566 2381}%
\special{fp}%
% VECTOR 2 0 3 0
% 2 2742 2781 2918 2781
% 
\special{pn 8}%
\special{pa 2742 2381}%
\special{pa 2918 2381}%
\special{fp}%
\special{sh 1}%
\special{pa 2918 2381}%
\special{pa 2851 2361}%
\special{pa 2865 2381}%
\special{pa 2851 2401}%
\special{pa 2918 2381}%
\special{fp}%
% VECTOR 2 0 3 0
% 2 4262 2613 4262 2445
% 
\special{pn 8}%
\special{pa 4262 2213}%
\special{pa 4262 2045}%
\special{fp}%
\special{sh 1}%
\special{pa 4262 2045}%
\special{pa 4242 2112}%
\special{pa 4262 2098}%
\special{pa 4282 2112}%
\special{pa 4262 2045}%
\special{fp}%
% VECTOR 2 0 3 0
% 2 2918 2285 2742 2285
% 
\special{pn 8}%
\special{pa 2918 1885}%
\special{pa 2742 1885}%
\special{fp}%
\special{sh 1}%
\special{pa 2742 1885}%
\special{pa 2809 1905}%
\special{pa 2795 1885}%
\special{pa 2809 1865}%
\special{pa 2742 1885}%
\special{fp}%
% VECTOR 2 0 3 0
% 2 1566 2445 1566 2613
% 
\special{pn 8}%
\special{pa 1566 2045}%
\special{pa 1566 2213}%
\special{fp}%
\special{sh 1}%
\special{pa 1566 2213}%
\special{pa 1586 2146}%
\special{pa 1566 2160}%
\special{pa 1546 2146}%
\special{pa 1566 2213}%
\special{fp}%
% STR 2 0 3 0
% 3 2406 681 2406 765 2 0
% $Re p_1$
\put(24.0600,-3.6500){\makebox(0,0)[lb]{$Im p_1$}}%
% STR 2 0 3 0
% 3 4749 2781 4749 2865 1 0
% $Im p_1$
\put(47.4900,-24.6500){\makebox(0,0)[lt]{$Re p_1$}}%
% STR 2 0 3 0
% 3 2204 2159 2204 2243 3 0
% $r_1$
\put(22.0400,-18.4300){\makebox(0,0)[rb]{$r_1$}}%
% VECTOR 2 0 3 0
% 2 2310 3020 2310 790
% 
\special{pn 8}%
\special{pa 2310 2620}%
\special{pa 2310 390}%
\special{fp}%
\special{sh 1}%
\special{pa 2310 390}%
\special{pa 2290 457}%
\special{pa 2310 443}%
\special{pa 2330 457}%
\special{pa 2310 390}%
\special{fp}%
% LINE 1 0 3 0
% 2 2890 2140 2990 2240
% 
\special{pn 13}%
\special{pa 2890 1740}%
\special{pa 2990 1840}%
\special{fp}%
% LINE 1 0 3 0
% 2 2980 2140 2890 2240
% 
\special{pn 13}%
\special{pa 2980 1740}%
\special{pa 2890 1840}%
\special{fp}%
% LINE 1 0 3 0
% 2 1600 2130 1700 2230
% 
\special{pn 13}%
\special{pa 1600 1730}%
\special{pa 1700 1830}%
\special{fp}%
% LINE 1 0 3 0
% 2 1690 2130 1600 2230
% 
\special{pn 13}%
\special{pa 1690 1730}%
\special{pa 1600 1830}%
\special{fp}%
\end{picture}%
%\epsfxsize=80mm
%\epsfysize=80mm
%\epsffile{floclityfi2.eps}
%\centerline{\epsfbox{ floclityfi.eps }}
\label{fig}
\caption{We can shift the integration path of $p_1$ upto the 
singularity closest to the real axis. Crosses in the figure 
represent the singularities.
The line integrals along the vertical lines 
at $\Re p_{1}=3\pi/2$ and $\Re p_{1}=-\pi/2$ cancel.
}
\end{center}
\end{figure}
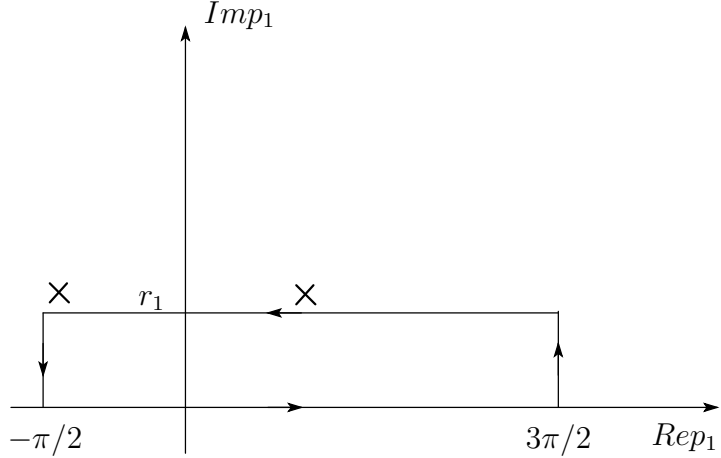

This property is also confirmed directly as follows:
We first recall the Riemann-Lebesgue lemma for a regular 
$H(p_{\mu})$
\begin{equation}
\lim_{x_{1}\rightarrow\infty}
D(x^{\mu},0)=\lim_{x_{1}\rightarrow\infty}\frac{1}{a}
\int_{-\pi/2}^{3\pi/2} \gamma_{5}H(p_{\mu})
\exp[ip_{\mu}x^{\mu}/a]d^{4}p_{\mu}/a^{4}=0
\end{equation}
and consider the case where the singularity 
closest to the real axis is written with real $q_{1}$ and 
positive $r_{1}$ as 
\begin{equation}
p_{1}=q_{1}+ir_{1}.
\end{equation}
We then consider the integration path parallel to the real
axis given by
\begin{equation}
p_{1}=q_{1}+i(r_{1}-\epsilon)
\end{equation}
where $\epsilon$ is a small positive number. The above integral
is then written as 
\begin{eqnarray}
D(x^{\mu},0)&=&\exp[-(r_{1}-\epsilon)x^{1}/a]\\
&&\times\frac{1}{a}
\int_{-\pi/2}^{3\pi/2} \gamma_{5}H(p_{\mu})
\exp[iq_{1}x^{1}/a+\sum^{\prime}_{\mu}ip_{\mu}x^{\mu}/a]
dq_{1}d^{3}p_{\mu}/a^{4}.\nonumber
\end{eqnarray}
Since the integrand is still regular, one can apply the 
Riemann-Lebesgue lemma to this last integral. This means
the exponential decay of the operator with 
(a conservative estimate of) the localization length $L$ given by
\begin{equation}
L=a/(r_{1}-\epsilon).
\end{equation}
 The existence of the well-defined mass gap thus  
leads to the exponential decay of the operator for a large spatial
separation. If the singularity of $H(p)$ in $p_{1}$, which is 
closest to the real axis, is written as 
$p_{1}=q_{1}+ir_{1}$, the above analysis gives an estimate of 
the localization length $L$ at
\begin{equation}
L\simeq a/r_{1}.
\end{equation}
Moreover, this estimate of the localization length is the 
same for  $H$ and for $H^{2k+1}=H_{(2k+1)}$, since we have shown
that the 
singularities in the complex $p_{1}$ (or $c_{1}$) plane are 
identical for  these two operators.

In the one-dimensional integral, one has
\begin{equation} 
\int^{\infty}_{-\infty}dy\exp[-|x-y|/L]\exp[-|y-z|/L]=
(L+|x-z|)\exp[-|x-z|/L]
\end{equation}
This shows that a multiplication of 
two operators, which decay exponentially, produces an operator 
which decays with the same exponential factor up to a 
polynomial prefactor. A generalization 
of this relation suggests that a suitable $2k+1$-th root of an 
exponentially decaying operator gives rise to an operator with 
an identical localization length for any finite $k$. This is 
what our analysis
of the free fermion operator so far indicates.
Later, we establish a finite locality domain of background
gauge fields for the operator $H^{2k+1}$ for any finite $k$.
Although we cannot explicitly demonstrate the locality domain
of $H$ in the interacting case, we infer the locality domain of 
$H$ from the above analysis of free fermion operators and a 
generalization of (5.13).

Incidentally, if the operator $H(p_{\mu})$ is analytic in the 
entire strip in the complex plane, i.e., if $r_{1}=\infty$ in 
(5.8), the localization length vanishes and the operator becomes 
an ultra-local one. 

\subsection{Crude Estimate of Localization Parameters}

We here present an estimate of the position of singularities
in  $H$, which is closest to the real axis, 
and illustrate an estimate of the localization length. 
The purposes of the analysis of a simplified version in this 
subsection are two-fold; firstly, we illustrate the explicit 
evaluation of the localization length, and secondly, we show that
the localization length can generally be much shorter than the 
estimate given by the general Legendre expansion in [9].
We examine the position of zeroes in 
\begin{eqnarray} 
f_{k}(c_{1})&\equiv&[1-c^{2}_{1}]^{2k+1}
+\{[1-c_{1}]^{2k+1}-1\}^{2}\nonumber\\
&=&[1-c_{1}]^{2(2k+1)}+1+[1-c^{2}_{1}]^{2k+1}
-2[1-c_{1}]^{2k+1}=0
\end{eqnarray}
This corresponds to the simplest $d=1$ dimensional example,
which is obtained from the general formula by setting 
$c^{\prime}_{\mu}=1$ ($\mu\neq 1$).
Since the values of the  variables $c^{\prime}_{\mu}=1$ 
($\mu\neq 1$) in (5.14) give the stationary point of the 
integrand of the 
Fourier transform of our problem with  $x^{\prime}_{\mu}=0$ 
($\mu\neq 1$), this 
example may also have some implications on the case $d>1$ for 
$k\gg 1$: The stationary
points have a non-vanishing phase space for the Fourier integral 
and thus they are expected to give an important contribution. 

The singularity in $c_{1}$ closest to the real axis 
 gives a dominant contribution for large $x_{1}$.
We thus look for the solution of 
\begin{equation}
f_{k}(z)=z^{2(2k+1)}+1+z^{2k+1}[(2-z)^{2k+1}-2]=0
\end{equation}
where we set $z=1-c_{1}$. 

The simplest case $k=0$ gives rise to\footnote{If one chooses
$2m_{0}=1$, $f_{0}(z)=5/4-c_{1}$ and thus $v_{max}=9/4, \ \ 
v_{min}=1/4$. This gives $\cosh\theta=\frac{(9/4+1/4)}
{(9/4-1/4)}=5/4$ and thus again agrees with the estimate on the 
basis of analyticity, $5/4-c_{1}=0$. This example shows that an
ultra-local operator and an exponentially local operator are 
located very close to each other.} 
\begin{equation}
f_{0}(z)=1
\end{equation}
and thus no singularity with $r_{1}=\infty$, namely an 
ultra-local operator. This estimate agrees with the one on the 
basis of the Legendre expansion
\begin{equation}
v_{max}=v_{min}=1
\end{equation}
and thus 
\begin{equation}
\cosh\theta=\frac{v_{max}+v_{min}}{v_{max}-v_{min}}=\infty.
\end{equation}

For $k\gg 1$, we assume that $z^{2k+1}$ is small 
for the solution. Then 
\begin{eqnarray}
&& 1+z^{2k+1}2^{2k+1}\simeq 0,\nonumber\\
&&z\simeq (-1)^{\frac{1}{2k+1}}\frac{1}{2}
=\exp[\pm\frac{i\pi}{2k+1}]\frac{1}{2} 
\end{eqnarray}
where we chose the solutions closest to the real axis in terms of 
$p_{1}$ defined by $c_{1}=\cos p_{1}$.
This value gives 
\begin{equation}
|z^{2k+1}|\simeq |\frac{1}{2^{2k+1}}|\ll 1
\end{equation}
and also a large value for  $(2-z)^{2k+1}$. Thus our estimate 
of the solution is consistent for large $k$.

We thus have
\begin{equation}
1-c_{1}\simeq\frac{1}{2}(\cos\frac{\pi}{2k+1}
\pm i\sin\frac{\pi}{2k+1}).
\end{equation}
If one sets $c_{1}=\cos(\theta_{0}+i\delta)$ with a 
{\em positive} $\delta$ and real $\theta_{0}$, we have
\begin{eqnarray}
c_{1}&=& \cos(\theta_{0}+i\delta)\nonumber\\
&=&\cos\theta_{0}\cosh\delta-i\sin\theta_{0}\sinh\delta
\nonumber\\
&=&1-\frac{1}{2}\cos\frac{\pi}{2k+1}
\mp \frac{i}{2}\sin\frac{\pi}{2k+1}
\end{eqnarray}
Since our equation (5.14) has real coefficients, we generally 
have a pair of solutions 
\begin{equation}
(\theta_{0}, \delta), \ \ \ (-\theta_{0}, \delta)
\end{equation}
where $\theta_{0}$ may be chosen to be positive; we thus 
explicitly solve only the positive solution in the following.
For large $k$ and small $\delta$, we have 
\begin{eqnarray}
&&\cos\theta_{0}\simeq 1-\frac{1}{2}\cos\frac{\pi}{2k+1},
\nonumber\\
&&\sin\theta_{0}\simeq \sqrt{1-\{1-\frac{1}{2}
\cos\frac{\pi}{2k+1}\}^{2}}\nonumber\\
&&\rightarrow \frac{\sqrt{3}}{2}.
\end{eqnarray}
From the relation
\begin{equation}
\sin\theta_{0}\sinh\delta\simeq \frac{1}{2}\sin\frac{\pi}{2k+1}
\end{equation}
we have
\begin{equation}
\delta\simeq \frac{1}{2\sin\theta_{0}}
\sin\frac{\pi}{2k+1}\simeq \frac{\pi}{\sqrt{3}(2k+1)}.
\end{equation}

From the above analysis of the Fourier transformation in (5.12), 
the  crude estimate of the localization length, which is defined 
by
\begin{equation}
|D(x_{1},0)|\sim \exp[-x^{1}\delta/a]=\exp[-x_{1}/L],
\end{equation}
is then given by 
\begin{eqnarray}
L/a&\simeq& \frac{\sqrt{3}}{\pi}(2k+1).
\end{eqnarray}

Generally, two (or more) solutions of 
$c_{1}=\cos(\theta_{0}+i\delta)$ with
an equal imaginary part and different real parts 
super-impose the oscillating behavior on 
the exponential decay with a period of the order of 
the lattice spacing[19].
For the case $k=0$ (the overlap operator ), the closest 
singularity generally appears on the imaginary axis and no 
oscillating behavior occurs.

We also performed a numerical analysis of the values $\theta_{0}
+i\delta$ in (5.22) and the localization
length by solving the equation (5.14) numerically for 
$k=1,3,5,7,9$. (See Table 1, where only the positive $\theta_{0}$
is shown). 
\begin{table}
\begin{center}
\begin{tabular}{lccr}
$k$ & $\theta_0$ & $\delta$ & $L/a$ \\ \hline
\it 1 & 1.69 & 0.298 & 3.35 \\
\it 3 & 1.62 & 0.118 & 8.47 \\
\it 5 & 1.60 & 0.0737 & 13.6 \\
\it 7 & 1.59 & 0.0535 & 18.7 \\
\it 9 & 1.59 & 0.0421 & 23.8 \\
\end{tabular}
\end{center}
\caption{Localization length for $k=1,3,5,7,9$}
\end{table}
Applying the method of least square fit  to these numerical data, 
we obtain the following relation
\begin{eqnarray}
L/a = 2.56 k + 0.8015.
\end{eqnarray}
This relation is not inconsistent with our crude analytical 
estimates (5.28).
We reiterate that there exist two solutions with an equal 
imaginary part and different real parts $\theta_0$ and 
$-\theta_0$
,which  explains the oscillating behavior super-imposed on the
exponential decay in 2-dimensional simulation[19]. 

Our estimates of the localization length (both of crude analytical
estimate and numerical estimate) are much smaller than 
the general estimate in Ref.[9], 
which gives for the present one-dimensional example of 
$H_{(2k+1)}$,
\begin{equation}
\cosh\theta=\frac{(2^{2k+1}-1)^{2}+1}{(2^{2k+1}-1)^{2}-1}
\simeq 1+2 2^{-2(2k+1)}
\end{equation}
and thus a much larger  localization length for $k\gg 1$
\begin{equation}
L/a\simeq (2k+1)2^{2k}.
\end{equation}
 
In the notation of (3.44), our estimate (5.19) is based on
\begin{equation}
f(\cos\theta)= 1+2^{2k+1}(1-\cos\theta)^{2k+1}\simeq 0
\end{equation}
which suggests that
\begin{eqnarray}
&&v_{min}=1,\nonumber\\
&&v_{max}=2^{2(2k+1)}
\end{eqnarray}
for large $k$. 
For $k\gg 1$, our function $f(\cos\theta)$ in (5.32) has a very 
small slope near 
$\cos\theta\simeq 1$, and thus the linear approximation gives a
poor estimate of the solution of $f(\cos\theta)=0$.
This particular property is not incorporated in the general
Legendre expansion.
In the general Legendre 
expansion, the huge maximum which appears in a specific sector 
of species doublers dominates the estimate of the 
localization length. We consider that this 
specific behavior of the integrand is the reason for the 
large discrepancy of our estimate from that of the general 
Legendre expansion [9]. The numerical simulation of the 
localization length in $d=2$ appears 
to support our estimate[19].

\section{Lower Bound for Interacting  
${D_{W}^{(2k+1)}}^{\dagger}D_{W}^{(2k+1)}$ }

In this section we evaluate the lower bound for 
${D_{W}^{(2k+1)}}^{\dagger}D_{W}^{(2k+1)}$ and establish a finite 
locality domain of background gauge fields for the 
operator $H^{2k+1}=H_{(2k+1)}$ for any finite $k$. This analysis
is a direct extension of the analyses in Refs.[9][10]. 
 
We first define the directional parallel transporters[10]:
\begin{eqnarray}
&&T_\mu(\psi)(x)= U_\mu(x)\psi(x+\hat\mu a)= \sum_{y}U_\mu(y)
\delta_{x+\hatmu a,y}\psi(y).
\end{eqnarray}
Note that $T_\mu$ are unitary operators, and their norms are $1$ 
due to the unitarity. Using $T_\mu$ and its Hermitian conjugate 
$T_\mu^\dagger$, $D_{W}^{(2k+1)}$ is written as
\begin{eqnarray}
D_{W}^{(2k+1)} &=& i\lc\f{1}{2}\gamma_\mu(T_\mu - T_\mu^\dagger)
\rc^{2k+1} + \lc\f{r}{2}\sum_{\mu}(1-T_\mu^\dagger)(1-T_\mu)
\rc^{2k+1}-\left(\f{m_0}{a}\right)^{2k+1}.
\end{eqnarray}
We then have 
\begin{eqnarray}
{D_{W}^{(2k+1)}}^{\dagger}D_{W}^{(2k+1)} &=& 
\lc\f{1}{2}\gamma_\mu(T_\mu - T_\mu^\dagger)\rc^{2(2k+1)}
+ \lc\f{1}{2}\sum_{\mu}(1-T_\mu^\dagger)(1-T_\mu)
\rc^{2(2k+1)}\nonumber\\
&&-2\lc\f{1}{2}\sum_{\mu}(1-T_\mu^\dagger)(1-T_\mu)
\rc^{2k+1} + 1\nonumber\\
&&-i\lc\f{1}{2}\gamma_\mu(T_\mu - T_\mu^\dagger)
\rc^{2k+1}\lc\f{1}{2}\sum_{\mu}
(1-T_\mu^\dagger)(1-T_\mu)\rc^{2k+1}\nonumber\\
&&+i\lc\f{1}{2}\sum_{\mu}(1-T_\mu^\dagger)(1-T_\mu)
\rc^{2k+1}\lc\f{1}{2}\gamma_\mu(T_\mu - T_\mu^\dagger)
\rc^{2k+1}.
\end{eqnarray}
Here we set $m_0=1,\, r=1$ and $a=1$ . Noting that the third term 
in the above equation is a non-positive operator, we rewrite 
the above equation as follows,
\begin{eqnarray}
{D_{W}^{(2k+1)}}^{\dagger}D_{W}^{(2k+1)} &=& 1+\lb\lc -\sum_{\mu}
\f{1}{4}(T_\mu - T_\mu^\dagger)^2\rc^{2k+1}\right.\nonumber\\
&&+\left.\lc \sum_{\mu}
\f{1}{4}(2-T_\mu-T_\mu^\dagger)^2\rc^{2k+1}-2\lc \sum_{\mu}
\f{1}{2}(1-T_\mu^\dagger)(1-T_\mu)\rc^{2k+1}\rb\nonumber\\
&&+\lb\lc -\sum_{\mu}
\f{1}{4}(T_\mu - T_\mu^\dagger)^2 + \f{1}{8}\sum_{\mu\neq\nu}
\gamma_\mu\gamma_\nu\lb T_\mu - T_\mu^\dagger,T_\nu - T_\nu^\dagger
\rb\rc^{2k+1}\right.\nonumber\\
&&\left.-\lc -\sum_{\mu}
\f{1}{4}(T_\mu - T_\mu^\dagger)^2\rc^{2k+1}\rb\nonumber\\
&&+\lb\lc\sum_{\mu}\f{1}{4}
(1-T_\mu^\dagger)(1-T_\mu)(1-T_\mu^\dagger)(1-T_\mu)\right.\right.
\nonumber\\
&& \left. +\sum_{\mu\neq\nu}
\f{1}{4}(1-T_\mu^\dagger)(1-T_\mu)(1-T_\nu^\dagger)(1-T_\nu)  
\rc^{2k+1}\nonumber\\
&&\left.-\lc\sum_{\mu}\f{1}{4}
(1-T_\mu^\dagger)(1-T_\mu)(1-T_\mu^\dagger)(1-T_\mu)\rc^{2k+1}\rb
\nonumber\\
&&+\lb -i\lc\f{1}{2}\gamma_\mu(T_\mu - T_\mu^\dagger)
\rc^{2k+1}\lc\f{1}{2}\sum_{\mu}
(1-T_\mu^\dagger)(1-T_\mu)\rc^{2k+1}\right.\nonumber\\
&&\left.+i\lc\f{1}{2}\sum_{\mu}(1-T_\mu^\dagger)(1-T_\mu)
\rc^{2k+1}\lc\f{1}{2}\gamma_\mu(T_\mu - T_\mu^\dagger)\rc^{2k+1}\rb.
\end{eqnarray}
From now on we refer to the expressions inside the  four large 
square brackets in (6.4) as $E_1, E_2, E_3, E_4$, respectively, 
and evaluate their norms in sequence.

\subsection{$E_1$}

We first define $A\equiv \sum_{\mu}\left((1-T_\mu)^2 + 
(1-T_\mu^\dagger)^2\right),\quad
B\equiv \sum_{\mu}(1-T_\mu^\dagger)(1-T_\mu)$.
Using $A$ and $B$, $E_1$ is written as
\begin{eqnarray}
E_1&\equiv&\lc -\sum_{\mu}
\f{1}{4}(T_\mu - T_\mu^\dagger)^2\rc^{2k+1}
+\lc \sum_{\mu}
\f{1}{4}(1-T_\mu^\dagger)(1-T_\mu)(1-T_\mu^\dagger)(1-T_\mu)
\rc^{2k+1}\nonumber\\
&&-2\lc \sum_{\mu}\f{1}{2}(1-T_\mu^\dagger)(1-T_\mu)\rc^{2k+1}
\nonumber\\
&=&\left(\f{1}{4}\right)^{2k+1}\lc (-A+2B)^{2k+1}+(A+2B)^{2k+1}\rc
-2\left(\f{1}{2}\right)^{2k+1}B^{2k+1}\nonumber\\
&=&\left(\f{1}{4}\right)^{2k+1}\lc 2(\underbrace{2A^{2k}B+\cdots
+2BA^{2k}}_{{}_{2k+1}C_1 })+2^3(\underbrace{2A^{2k-2}B^3 + 
\cdots +2B^3A^{2k-2}}_{{}_{2k+1}C_2 })\right.\nonumber\\
&&\left.+ \cdots +2^{2k-1}(\underbrace{2A^2B^{2k-1} + \cdots 
+2B^{2k-1}A^2}_{{}_{2k+1}C_1 })\rc.
\end{eqnarray} 
In the last expression of (6.5), the notation such as 
${}_{2k+1}C_1$ shows the 
number of terms.
We also note that only the even power of $A$ and the odd power of
$B$ appear in the last expression of (6.5). From this fact,
all operators 
appearing in the above equation can be rewritten as a sum of 
non-negative 
operators such as $2^{l+2}A^mB^{2l+1}A^m$ and the terms 
containing a commutator $[A,B]$. To show that  $2^{l+2}
A^mB^{2l+1}A^m$ are 
non-negative operators, we rewrite them as
\begin{eqnarray}
2^{l+2}A^mB^{2l+1}A^m &=& 2^{l+2}A^mB^l\left(\sum_{\mu}
(1-T_\mu^\dagger)(1-T_\mu)\right)B^lA^m\nonumber\\
&=& 2^{l+2}\sum_{\mu}
\left((1-T_\mu)B^lA^m\right)^\dagger\left((1-T_\mu)B^lA^m\right).
\end{eqnarray}
by noting $A^\dagger = A$ and 
$B^\dagger =B$.

Now in each term in the last expression of (6.5) we want to 
bring all $B$ to the center 
of the powers of $A$, but performing this
manipulation for all the operators is quite tedious.
To cope with this, we first note that
the commutators, which appear when all $B$ are brought to the 
center of the powers of $A$, are proportional to the field 
strength and that we only need the upper bound to those terms 
containing field strength when we finally evaluate the lower 
bound for ${D_{W}^{(2k+1)}}^{\dagger}D_{W}^{(2k+1)}$. We 
therefore evaluate
$E_1$ approximately as the number of terms multiplied to the 
term which  produces the 
largest number of terms containing the commutator. (By this way,
we estimate the coefficient of the right-hand side of (6.7) below
at a value larger than the true value. But this is a safe 
operation in the estimate of the upper bound.) 
The terms, which produce the largest number of terms containing
the commutator, are the terms 
that include $B$ at the right-most side as in 
$A^{2k-2l}B^{2l+1}$, and we obtain 
 $(k-l)(2l+1)$ terms containing the  commutator.

As an example, we evaluate the norm of the term containing the 
$(2l+1)-$th 
power of $B$. Using the triangle inequality and 
 $\| AB\|\le \| A\|\| B\|$, we obtain
\begin{eqnarray}
&&\| 2^{2l+2}(\underbrace{A^{2k-2l}B^{2l+1} + \cdots + 
B^{2l+1}A^{2k-2l}}_{{}_{2k+1}C_{2l+1} })-(\underbrace{2^{2l+2}
{}_{2k+1}C_{2l+1}A^{k-l}B^{2l+1}A^{k-l}}_{non-negative\, 
operators})\|
\nonumber\\
&&\leq 2^{2l+2}{}_{2k+1}C_{2l+1}(k-l)(2l+1)\| A\|^{2k-2l-1}
\| B\|^{2l}\| [A,B]\|,\qquad l=0,1,\cdots k-1.
\end{eqnarray}
Now we impose the constraint on the commutators of $T_\mu$ and 
$T_\mu^\dagger$ as follows
\begin{eqnarray}
\| [T_\mu,T_\nu]\|=\| [T_\mu,T_\nu^\dagger]\|
=\| [T_\mu^\dagger,T_\nu^\dagger]\|= \| 1-U_{\mu\nu}\| < 
\epsilon,\quad \mu\neq\nu.
\end{eqnarray}
Then the norms of $\| A\|$, $\|B\|$ and $\|[A,B]\|$ are 
evaluated by using triangle inequality as 
\begin{eqnarray}
\| A\| \le 32,\quad \| B\| \le 16,\quad \| [A,B] \| < 96\epsilon.
\end{eqnarray}
We thus have 
\begin{eqnarray}
&&\| E_1 - \left(\f{1}{4}\right)^{2k+1}
\underbrace{\sum_{l=0}^{k-1}(2^{2l+2}{}_{2k+1}C_{2l+1}
A^{k-l}B^{2l+1}A^{k-l})}_{non-negative\, 
operators}\|
\nonumber\\
&&\le \left(\f{1}{4}\right)^{2k+1}\sum_{l=0}^{k-1}2^{2l+2}
{}_{2k+1}C_{2l+1}(k-l)(2l+1)\| A\|^{2k-2l-1}
\| B\|^{2l}\| [A,B]\| \nonumber\\
&& < 3\sum_{l=0}^{k-1}
{}_{2k+1}C_{2l+1}(k-l)(2l+1)64^k\epsilon.
\end{eqnarray}

\subsection{$E_2$}

Here we define $A\equiv -\sum_{\mu}
\f{1}{4}(T_\mu - T_\mu^\dagger)^2,\quad B\equiv \f{1}{8}
\sum_{\mu\neq\nu}
\gamma_\mu\gamma_\nu\lb T_\mu - T_\mu^\dagger,T_\nu - T_\nu^\dagger
\rb$. Since all the terms in
\begin{equation}
E_2\equiv (A+B)^{2k+1}-A^{2k+1}
\end{equation}
contain $B$ and the norm of $B$ is 
bounded by a 
value proportional to $\epsilon$, we can evaluate the upper bound
to the norm of $E_2$. As in the previous subsection,
noting that $\| A\|\le 4$ and $\| B\|\le 6\epsilon$, we evaluate
\begin{eqnarray}
\| E_2\|\equiv\|(A+B)^{2k+1}-A^{2k+1}\|&\le&\left(\| A\| + 
\| B\|\right)^{2k+1} - \| A\|^{2k+1}\nonumber\\
&<&\sum_{l=1}^{2k+1}{}_{2k+1}C_{l}4^{2k+1-l}(6\epsilon)^{l}.
\end{eqnarray}

\subsection{$E_3$}

We first rewrite $E_3$ as follows,
\begin{eqnarray}
E_3&\equiv&\lc\sum_{\mu}\f{1}{4}
(2-T_\mu - T_\mu^\dagger)^2 +\sum_{\mu\neq\nu}
\f{1}{8}\left((1-T_\mu)(1-T_\nu^\dagger)(1-T_\nu^\dagger)
(1-T_\mu^\dagger)\right.\right.\nonumber\\
&&\left.\left. + (1-T_\mu^\dagger)(1-T_\nu)(1-T_\nu^\dagger)
(1-T_\mu)\right)
-\sum_{\mu\neq\nu}\f{1}{8}\left(T_\mu[T_\mu^\dagger,T_\nu 
+T_\nu^\dagger] + T_\mu^\dagger[T_\mu,T_\nu + T_\nu^\dagger]\right)
\rc^{2k+1}\nonumber\\
&& - \lc\sum_{\mu}\f{1}{4}(2-T_\mu - T_\mu^\dagger)^2
\rc^{2k+1}
\nonumber\\
&=&(A+B+C)^{2k+1}-A^{2k+1}\nonumber\\
&=&\lc \underbrace{A^{2k}(B+C)+\cdots +(B+C)A^{2k}}_{
{}_{2k+1}C_{1}}\rc 
+\lc \underbrace{A^{2k-1}(B+C)^2+\cdots +(B+C)^2A^{2k-1}}_{
{}_{2k+1}C_{2}}\rc\nonumber\\
&&+\cdots + \lc \underbrace{A(B+C)^{2k}+\cdots +(B+C)^{2k}A}_{
{}_{2k+1}C_{1}}\rc + (B+C)^{2k+1},
\end{eqnarray}
where
\begin{eqnarray}
A&\equiv&\sum_{\mu}\f{1}{4}
(2-T_\mu-T_\mu^\dagger)^2,\\
B&\equiv& \sum_{\mu\neq\nu}
\f{1}{8}\{(1-T_\mu)(1-T_\nu^\dagger)(1-T_\nu^\dagger)
(1-T_\mu^\dagger)\nonumber\\
&&+ (1-T_\mu^\dagger)(1-T_\nu)(1-T_\nu^\dagger)
(1-T_\mu)\},\\
C&\equiv& -\sum_{\mu\neq\nu}\f{1}{8}
\left(T_\mu[T_\mu^\dagger,T_\nu 
+T_\nu^\dagger] + T_\mu^\dagger[T_\mu,T_\nu 
+ T_\nu^\dagger]\right).
\end{eqnarray}
Here we note that $A$ and $B$ are non-negative and hermitian  
operators and that the norm of $C$ is bounded by a value proportional 
to $\epsilon$. We split $E_3$ into the terms containing $C$ 
and the terms not containing $C$, and evaluate their norms 
in sequence.

\subsubsection{Terms Not Containing $C$ }

We further split those terms, which do not contain $C$,
 into the 
terms with an even power of $A$ and the terms with an 
odd power of $A$. We first evaluate the terms with an
even power of $A$. We then examine
\begin{eqnarray}
&&\sum_{l=0}^{k}(\underbrace{A^{2k-2l}B^{2l+1}+ \cdots 
+ B^{2l+1}A^{2k-2l}}_{{}_{2k+1}C_{2l+1}}).
\end{eqnarray}
Since  $A^{k-l}B^{2l+1}A^{k-l}$  is a non-negative operator,
we want to write all the terms as a sum of those non-negative 
operators and the terms containing $[A,B]$. Here we adopt 
the same procedure as used in $E_1$ above. The terms, which 
produce the largest number of terms containing the commutator,
 are the terms that contain $B$ at the right-most side 
as in $A^{2k-2l}B^{2l+1}$. All the terms in (6.17) put together 
produce $(k-l)(2l+1)$ terms which contain the commutator. 
We thus have
\begin{eqnarray}
&&\| A^{2k-2l}B^{2l+1}+ \cdots + B^{2l+1}A^{2k-2l}
- \underbrace{{}_{2k+1}C_{2l+1}A^{k-l}B^{2l+1}A^{k-l}}_{
non-negative\, operators} \|\nonumber\\
&& < {}_{2k+1}C_{2l+1}(k-l)(2l+1)\| A\|^{2k-2l-1} \| B\|^{2l}
\| [A,B]\|. 
\end{eqnarray}
Noting that the norms of $A$, $B$ and $[A,B]$ are bounded as 
$\| A\|\le 16,\quad
\| B\|\le 48,\quad \| [A,B]\|\le 192\epsilon$, respectively,
we obtain
\begin{eqnarray}
&&\|\sum_{l=0}^{k}( A^{2k-2l}B^{2l+1}+ \cdots + B^{2l+1}A^{2k-2l})
-\sum_{l=0}^{k} 
\underbrace{{}_{2k+1}C_{2l+1}A^{k-l}B^{2l+1}A^{k-l}}_{
non-negative\, operators} \|\nonumber\\
&&<\sum_{l=0}^{k} {}_{2k+1}C_{2l+1}(k-l)(2l+1) 16^{2k-2l-1} 
48^{2l}192\epsilon. 
\end{eqnarray}

Next we examine the terms with an odd power of $A$
\begin{eqnarray}
&&\sum_{l=1}^{k}(\underbrace{A^{2k-2l+1}B^{2l}+ 
\cdots + B^{2l}A^{2k-2l+1}}_{{}_{2k+1}C_{2l}}).
\end{eqnarray}
We use the same procedure as above. Since $B^lA^{2k-2l+1}B^l$ is 
a non-negative operator and the terms, which produce the largest 
number of terms with the commutator, are the ones 
that contain $B$ at the right-most side, we obtain after counting
all the terms in (6.20) 
\begin{eqnarray}
&&\|\sum_{l=1}^{k}(A^{2k-2l+1}B^{2l}+ 
\cdots + B^{2l}A^{2k-2l+1})- \sum_{l=1}^{k}{}_{2k+1}C_{2l}
B^lA^{2k-2l+1}B^l\|\nonumber\\
&&<  \sum_{l=1}^{k}{}_{2k+1}C_{2l}l(2k-2l+1)
\| A\|^{2k-2l}\| B\|^{2l-1}\| [A,B]\|\nonumber\\
&&<\sum_{l=1}^{k}{}_{2k+1}C_{2l}l(2k-2l+1)
16^{2k-2l}48^{2l-1}192\epsilon.
\end{eqnarray}

\subsubsection{Terms Containing $C$ }

Noting that the norm of $C$ is bounded as $\| C\|\le 6\epsilon$,
one can confirm that the norm of the terms 
containing $C$ is bounded from above by
\begin{eqnarray}
&&\sum_{m=1}^{2k+1}{}_{2k+1}C_{m}\| A\|^{2k+1-m}
\sum_{l=0}^{m-1}{}_{m}C_{l}\| B\|^l\| C\|^{m-l}\nonumber\\
&&<\sum_{m=1}^{2k+1}\sum_{l=0}^{m-1}{}_{2k+1}C_{m}{}_{m}C_{l}
16^{2k+1-m}48^l(6\epsilon)^{m-l}. 
\end{eqnarray}

\subsubsection{Final Result of  $E_{3}$ }

Collecting all the caluculations in this subsection, we have  
\begin{eqnarray}
&&\| E_3 -\left(\underbrace{\sum_{l=0}^{k} 
{}_{2k+1}C_{2l+1}A^{k-l}B^{2l+1}A^{k-l} + \sum_{l=1}^{k}
{}_{2k+1}C_{2l}
B^lA^{2k-2l+1}B^l}_{non-negative\,operators}\right)\|\nonumber\\
&&< 2\sum_{l=1}^{2k+1} {}_{2k+1}C_{l}\,l(2k+1-l) 16^{2k-l} 
48^{l}\epsilon\nonumber\\
&&\quad + \sum_{m=1}^{2k+1}\sum_{l=0}^{m-1}{}_{2k+1}C_{m}{}_{m}C_{l}
16^{2k+1-m}48^l(6\epsilon)^{m-l}. 
\end{eqnarray}

\subsection{$E_4$}

We write $E_4$ as 
\begin{eqnarray}
E_4 &\equiv& -iA^{2k+1}B^{2k+1} +iB^{2k+1}A^{2k+1}, 
\end{eqnarray}
where
\begin{eqnarray}
A&\equiv&\f{1}{2}\gamma_\mu(T_\mu - T_\mu^\dagger),\nonumber\\
B&\equiv&\f{1}{2}\sum_{\mu}(2-T_\mu-T_\mu^\dagger).
\end{eqnarray}
The commutator $[A^{2k+1},B^{2k+1}]$ is written as a sum of 
$(2k+1)^2$ terms containing the commutator 
$[A,B]$. We then have
\begin{eqnarray}
\|-iA^{2k+1}B^{2k+1} +iB^{2k+1}A^{2k+1} \|\le (2k+1)^2
\| A\|^{2k}\| B\|^{2k}\| [A,B]\|. 
\end{eqnarray}
Noting that the norms of $A$, $B$ and $[A,B]$ are bounded as 
$\| A\|\le 4,\quad \| B\|\le 8,\quad 
\| [A,B]\|< 6\sqrt{2}\epsilon$, 
respectively\footnote{Note that $[A,B]$ can be rewritten as
\begin{eqnarray}
[A,B]&=&[\f{1}{2}\gamma_\mu(T_\mu - T_\mu^\dagger),
\f{1}{2}\sum_{\mu}(2-T_\mu-T_\mu^\dagger)]\nonumber\\
&=&-\f{1}{8}\sum_{\mu\neq\nu}\lc(\gamma_\mu-\gamma_\nu)
([T_\mu,T_\nu]-[T_\mu^\dagger,T_\nu^\dagger])+
(\gamma_\mu+\gamma_\nu)([T_\mu,T_\nu^\dagger]-
[T_\mu^\dagger,T_\nu])\rc,\nonumber
\end{eqnarray}
and that $(\gamma_\mu\pm\gamma_\nu)
(\gamma_\mu\pm\gamma_\nu)^\dagger = 2$ for $\mu\neq\nu$.}, we obtain
\begin{eqnarray}
&&\| E_4\| < 6\sqrt{2}(2k+1)^2 16^k 64^{k}\epsilon.  
\end{eqnarray}

\subsection{Lower Bound for Interacting
${D_{W}^{(2k+1)}}^{\dagger}D_{W}^{(2k+1)}$}

Finally, we combine all the  caluculations in this section and 
obtain the lower bound for interacting 
${D_{W}^{(2k+1)}}^{\dagger}D_{W}^{(2k+1)}$ as
\begin{eqnarray}
\|{D_w^{(2k+1)}}^{\dagger}D_w^{(2k+1)}\|&=&\| 1+E_1+ E_2 +E_3 +E_4 \|
\nonumber\\
&\ge&\| 1+ \underbrace{\left(\f{1}{4}\right)^{2k+1}
\sum_{l=0}^{k-1}(2^{2l+2}{}_{2k+1}C_{2l+1}
A^{k-l}B^{2l+1}A^{k-l})}_{non-negative\, operators}\nonumber\\
&&+\underbrace{\sum_{l=0}^{k} 
{}_{2k+1}C_{2l+1}A^{k-l}B^{2l+1}A^{k-l} + \sum_{l=1}^{k}
{}_{2k+1}C_{2l}
B^lA^{2k-2l+1}B^l}_{non-negative\, operators}\|\nonumber\\
&&-\| E_1 - \left(\f{1}{4}\right)^{2k+1}
\sum_{l=0}^{k-1}(2^{2l+2}{}_{2k+1}C_{2l+1}
A^{k-l}B^{2l+1}A^{k-l})\|\nonumber\\
&& -\|E_2\| \nonumber\\
&&-\| E_3 -\left(\sum_{l=0}^{k} 
{}_{2k+1}C_{2l+1}A^{k-l}B^{2l+1}A^{k-l} + \sum_{l=1}^{k}
{}_{2k+1}C_{2l}
B^lA^{2k-2l+1}B^l\right)\|\nonumber\\
&&- \| E_4\|\nonumber\\
&>& 1 - \lb
3\sum_{l=0}^{k-1}
{}_{2k+1}C_{2l+1}(k-l)(2l+1)64^k\epsilon\right. \nonumber\\
&&\quad +\sum_{l=1}^{2k+1}{}_{2k+1}C_{l}4^{2k+1-l}(6\epsilon)^{l}
\nonumber\\
&&\quad + 2\sum_{l=1}^{2k+1} {}_{2k+1}C_{l}\,l(2k+1-l) 
16^{2k-l} 
48^{l}\epsilon\nonumber\\
&&\quad + \sum_{m=1}^{2k+1}\sum_{l=0}^{m-1}{}_{2k+1}C_{m}{}_{m}C_{l}
16^{2k+1-m}48^l(6\epsilon)^{m-l}\nonumber\\
&&\left.\quad + 6\sqrt{2}(2k+1)^2 16^k 64^{k}\epsilon\rb.
\end{eqnarray} 
By taking $k=0$, for example, in the above inequality, we have 
\begin{eqnarray}
\|{D_{W}}^{\dagger}D_{W}\|&>& 1 - 6(2+\sqrt{2})\epsilon.
\end{eqnarray}
This inequality naturaly agrees with that derived by 
Neuberger[10] for the Wilson-Dirac operator, which corresponds 
to $k=0$.
The original result of Hernandez, Jansen and  L\"{u}scher[9]
gives a lower bound $1-30\epsilon$. 

Taking $k=1$,for example,  we obtain approximately
\begin{eqnarray}
\|{D_{W}^{(3)}}^{\dagger}D_{W}^{(3)}\|&>& 1 - 2\times10^5\epsilon.
\end{eqnarray}
It is observed that the present estimate of the locality domain 
of gauge field strength for 
$H_{(2k+1)}$ gives a  very small value. However, we have  
set all the non-negative operators in 
$\|{D_{W}^{(3)}}^{\dagger}D_{W}^{(3)}\| $ at $0$ in our estimate,
as in (6.28), and those terms could be just as 
large as the coefficients of $\epsilon$, which are estimated by 
conservative huge upper bounds. We thus expect that  
the actual locality domain of 
gauge field strength could be much larger.

\subsection{Upper Bound for 
${D_{W}^{(2k+1)}}^{\dagger}D_{W}^{(2k+1)}$}
We next evaluate the upper bound for 
${D_{W}^{(2k+1)}}^{\dagger}D_{W}^{(2k+1)}$. Using the triangle
inequality and $\| AB\|\le \| A \| \| B \|$, we can evaluate the
upper bound for $D_{W}^{(2k+1)}$
as,
\begin{eqnarray}
\| D_{W}^{(2k+1)}\|&\le& 1+ 4^{2k+1} + 8^{2k+1}, 
\end{eqnarray}
and then the upper bound for ${D_{W}^{(2k+1)}}^{\dagger}D_{W}^{(2k+1)}$
is evaluated as
\begin{eqnarray}
\| {D_{W}^{(2k+1)}}^{\dagger}D_{W}^{(2k+1)}\| &\le& 
\left(1+ 4^{2k+1} + 8^{2k+1}\right)^2.
\end{eqnarray}

\subsection{Locality of Interacting $H=a\gamma_{5}D$}

Our analysis of ${D_{W}^{(2k+1)}}^{\dagger}D_{W}^{(2k+1)}$
establishes the locality domain of the operator 
\begin{equation}
H_{(2k+1)}\equiv(\gamma_{5}aD)^{2k+1}=\frac{1}{2}\gamma_{5}
[1+D_{W}^{(2k+1)}\frac{1}
{\sqrt{(D_{W}^{(2k+1)})^{\dag}D_{W}^{(2k+1)}}}].
\end{equation}
for any finite $k$. Since this operator appears in the generalized
chiral operator
\begin{equation}
\Gamma_{5}=\gamma_{5}-H_{(2k+1)}
\end{equation}
our analysis provides a basis for the cohomological analyses
of chiral gauge theory such as in [15][16].
Note that $\Gamma_{5}$ defines the index 
$Tr\Gamma_{5}=n_{+}-n_{-}$. Our analysis in this section, which 
is regarded as an extension of chiral anomaly calculation,
is naturally consistent with the analysis of anomaly in [13].

We have not established the locality domain of gauge fields for 
the operator
\begin{equation}
H=a\gamma_{5}D=(H_{(2k+1)})^{1/(2k+1)}
\end{equation}
explicitly in the interacting case. We however believe that 
$H$ is local with approximately the same locality domain 
of gauge fields as for $H_{(2k+1)}$. The reasons for this 
expectation are two-fold: The first is that we have established
the locality of free $H$ with a well-defined mass gap, 
which agrees with the mass gap for $H_{(2k+1)}$.
 The second reason is a generalization 
of (5.13). A multiplication of exponentially decaying operators
also produces an exponentially decaying operator with the same 
localization length up to a possible polynomial prefactor.
This suggests that the operation (6.35) also preserves the 
property of exponential decay for any finite $k$ in the 
interacting case also.

\section{Discussion and Conclusion}

We have discussed the locality properties of the general class of 
lattice Dirac operators defined by (1.1), which satisfy the 
index theorem for any finite $k$. We first established the 
locality of all 
these operators at the free fermion level. We also presented 
a crude estimate of the localization length for these operators:
The localization length approximately increases as $2k+1$,
which is much shorter than the estimate on the basis of general
Legendre expansion.
It is clear that the operator, which is local at the free fermion 
level with a well-defined
``mass gap'', is also local for sufficiently small and smooth
background gauge fields. 
We in fact established a finite locality domain of gauge fields
for $H_{(2k+1)}$ for any finite $k$ by extending the analyses 
in [9][10].
This locality is sufficient 
for the cohomological analyses of non-Abelian 
chiral gauge theory[15][16], for example; note that our 
operators satisfy the exact index theorem on the lattice. 
We also argued that the locality domain of gauge fields for 
$H$ itself is approximately the same as for $H_{(2k+1)}$.
From a view point of the regularization of continuum gauge theory,
our operators with any finite $k$ provide a satisfactory 
lattice regularization. However, our (conservative) estimate of 
the locality domain (6.28) is quite small. From a view point of 
numerical simulation, we thus need to perform a detailed numerical
study of the locality domain and see if the actual locality 
domain is much larger, as we naively expect. 
Once a larger locality domain is established by numerical
analyses, our general operators
may be used for an explicit construction of chiral theories
[15][16][20]-[24], for example. 

The present analysis of locality properties is also 
closely related to the 
perturbative evaluation of chiral $U(1)$ anomaly for 
$|agA_{\mu}|\ll 1$ with a fixed lattice spacing, and the 
independence of anomaly coefficient for a small variation of 
the parameters $r$ and $m_{0}$[13]. See the similar analyses of 
anomaly for the overlap operator[24], which corresponds to 
$k=0$.

As for the practical implications of our general operators,
these operators exhibit properties somewhat analogous 
to those of the ``perfect action'' of Hasenfratz et al.[26]; both
of these operators exhibit better chiral properties closer to
those of the continuum operator. Also the spectrum of the 
opeators with $k>0$ is closer to that of the continuum operator 
in the sense that 
the small eigenvalues of $D$ accumulate along  the 
imaginary axis[17] compared to the standard overlap operator, 
for which the eigenvalues of $D$ draw a circle in the complex 
eigenvalue plane. Contrary to an approximate solution to
the perfect action[26], the explicit construction of our 
operators satisfies the index theorem exactly for all finite $k$.
It may be interesting to apply our general operator with 
$k=1$, for example, to practical QCD simulations, since our 
operators preserve all the good chiral properties[27][28] of the 
overlap operator.\\

One of us (K.F.) thanks T-W. Chiu for helpful correspodences
about the numerical analyses, which were essential for the 
present study, and Y. Matsuo for a clarifying 
comment on the Fourier transform of an analytic function.


\begin{thebibliography}{1}
\bibitem{1}
P.H. Ginsparg and K.G. Wilson, Phys. Rev. {\bf D25} (1982)2649.
\bibitem{2}
H. Neuberger, Phys. Lett.{\bf B417}(1998)141;{\bf B427}(1998)353.
\bibitem{3} 
P. Hasenfratz, V. Laliena and F. Niedermayer, Phys. Lett. 
{\bf B427}(1998)125.
\bibitem{4}
M. L\"{u}scher, Phys. Lett. {\bf B428}(1998)342.
\bibitem{5}
D.~B.~Kaplan,
Phys.\ Lett.\ {\bf B288} (1992) 342.
\bibitem{6}
Y.~Shamir,
Nucl.\ Phys.\ {\bf B406} (1993) 90.\\
V.~Furman and Y.~Shamir,
Nucl.\ Phys.\ {\bf B439} (1995) 54.
\bibitem{7}
R.~Narayanan and H.~Neuberger,
Phys.\ Rev.\ Lett.\ {\bf 71} (1993) 3251.\\
R.~Narayanan and H.~Neuberger,
Nucl.\ Phys.\ {\bf B412} (1994) 574.
\bibitem{8}
P. Hasenfratz and F. Niedermayer, Nucl. Phys. {\bf B414}(1994)785.\\
T. DeGrand, A. Hasenfratz, P. Hasenfratz, P. Kunst and F. Niedermayer, 
Nucl.
Phys. {\bf B}(Proc. Suppl) 53 (1997) 942.\\
W. Bietenholtz and U. J. Wiese, Nucl. Phys. {\bf B464} (1996)319; Nucl.
Phys. {\bf B} (Proc. Suppl) 47 (1996) 575.
\bibitem{9}
P. Hernandez, K. Jansen and M. L\"{u}scher, Nucl. Phys. {\bf B552}
,363 (1999).
\bibitem{10}
H. Neuberger, Phys.Rev.{\bf D61}(2000)085015.
\bibitem{11}
I.~Horvath,
Phys.\ Rev.\ Lett.\ {\bf 81} (1998) 4063; Phys.\ Rev.\ 
{\bf D 60} (1999) 034510.
\bibitem{12}
K. Fujikawa, Nucl. Phys. {\bf B589}(2000)487.
\bibitem{13}
K. Fujikawa and M. Ishibashi, Nucl. Phys. {\bf B587}(2000)419.
\bibitem{14}
Y. Kikukawa,
Nucl.\ Phys.\ {\bf B584} (2000) 511.
\bibitem{15}
M. L\"{u}scher, Nucl. Phys. {\bf B568}(2000)162 ; JHEP {\bf 06}
(2000)028.
\bibitem{16}
H. Suzuki, Nucl. Phys.{\bf B585} (2000)471.\\
H.~Igarashi, K.~Okuyama and H.~Suzuki,
``Errata and addenda to 'Anomaly cancellation condition in 
lattice gauge theory','' hep-lat/0012018.
\bibitem{17}
T.W. Chiu, hep-lat/0010070.\\
See also  
T.W. Chiu, Nucl. Phys. {\bf B588} (2000) 400; hep-lat/0008010.
\bibitem{18}
P.M. Morse and H. Feshbach, {\em Methods in Theoretical Physics},
Chapter 4 (McGraw-Hill, New York, 1953).
\bibitem{19}
T.W. Chiu, in preparation.
\bibitem{20}
M. L\"{u}scher, Nucl.Phys.{\bf B549}(1999)295.
\bibitem{21}
H. Neuberger, Phys. Rev.{\bf D59}(1999)085006.\\
H.~Neuberger, Phys.\ Rev.\ {\bf D 63} (2001) 014503
\bibitem{22}
H. Suzuki, Prog. Theor. Phys. {\bf 101} (1999)1147.
\bibitem{23}
M. L\"{u}scher,
``Chiral gauge theories revisited,'' hep-th/0102028.
\bibitem{24}
M.~Golterman,
``Lattice chiral gauge theories,'' hep-lat/0011027.
\bibitem{25}
Y. Kikukawa and A. Yamada, Phys. Lett. {\bf B448}(1999)265.\\
D.H. Adams, hep-lat/9812003.\\
H. Suzuki, Prog.Theor.Phys.{\bf 102}(1999)141.\\
K. Fujikawa, Nucl.Phys.{\bf B546}(1999)480.
\bibitem{26}
P. Hasenfratz, Nucl. Phys. Proc. Suppl. {\bf 63} (1998) 53 ;
Nucl. Phys. {\bf B25} (1998) 401.\\
P. Hasenfratz, S. Hauswirth, K.Holland, T. J\"{o}rg, 
F. Niedermayer
and U. Wegner, hep-lat/0010061, and references therein. 
\bibitem{27}
S. Chandrasekharan, Phys. Rev. {\bf D60} (1999) 074503.
\bibitem{28}
F. Niedermayer, Nucl. Phys, Proc. Suppl.{\bf 73} (1999) 105, 
and references therein.
\end{thebibliography}
\end{document}